\documentclass[12pt]{article}
\usepackage{amsmath}
\usepackage{color}
\usepackage{graphicx,epsfig}
\usepackage{cite}
\title{Effects of nonlocal interactions on s- and d-wave superconducting correlations
in the extended Hubbard model}
\author{Pavol Farka\v sovsk\'y\\
Institute  of  Experimental  Physics,  Slovak   Academy   of
Sciences\\
Watsonova 47, 040 01 Ko\v {s}ice, Slovakia\\
E-mail: farky@saske.sk}
\date{}
\begin{document}
\baselineskip=22pt
\maketitle
\begin{abstract}
We investigate the influence of nonlocal interactions on superconducting
correlations within the extended Hubbard model. In addition to the on-site
Coulomb interaction and nearest-neighbor hopping, we include next-nearest-neighbor
hopping together with several physically relevant nonlocal terms, namely
nearest-neighbor Coulomb interaction, correlated hopping, exchange interaction,
and pair hopping. Using Lanczos exact diagonalization on a $4\times4$ cluster,
supported by projector quantum Monte Carlo simulations for selected parameter
regimes, we analyze pairing correlations in both the $s$- and $d$-wave channels.
We demonstrate that nonlocal interactions exert a highly nontrivial and
symmetry-dependent influence on superconducting correlations. While the on-site
repulsion in cooperation with next-nearest-neighbor hopping enhances $d$-wave
pairing tendencies, correlated hopping and nearest-neighbor Coulomb interaction
strongly promote $s$-wave correlations, whereas exchange and pair-hopping 
interactions can efficiently suppress
superconductivity beyond relatively small critical strengths. When all nonlocal
interactions are considered simultaneously, the resulting phase diagrams reveal
a complex interplay and competition between different pairing symmetries.
Our results highlight the crucial role of extended interactions in shaping the
pairing landscape of strongly correlated systems and demonstrate that a
comprehensive treatment beyond the minimal Hubbard model is essential for a
realistic description of unconventional superconductivity.
\end{abstract}

\newpage
\section{Introduction}
Understanding the role of strong electronic correlations in unconventional 
superconductivity remains one of the central open problems in condensed matter physics. 
In particular, the microscopic mechanisms underlying superconductivity in strongly correlated
materials, most notably the cuprate high-temperature superconductors, continue to motivate extensive theoretical 
studies~\cite{Scalapino,Wolf,Jiang,Roig}. A paradigmatic minimal model believed to capture the essential 
physics of these systems is the two-dimensional single-band Hubbard model~\cite{Hubbard}, which incorporates 
an on-site repulsive Coulomb interaction.

It is by now widely accepted that doping the Hubbard model away from half filling gives rise to a superconducting 
ground state with d-wave symmetry. This conclusion is supported by a large body of work employing 
a variety of analytical and numerical techniques. Nevertheless, controlled and exact results remain 
scarce. Exact solutions are limited to special cases, such as asymptotically weak coupling in the 
thermodynamic limit~\cite{Kohn,Raghu1}, while the physically relevant intermediate- to strong-coupling regime 
has primarily been explored using numerical methods, including determinant and auxiliary-field 
quantum Monte Carlo simulations~\cite{Maier1,Maier2,Haule}, calculations within the related
$t$–$J$ model~\cite{Anderson,Lee}, and exact 
diagonalization studies on finite clusters~\cite{Rigol1,Rigol2,Jia}.

Although the on-site Coulomb repulsion alone is sufficient to generate superconductivity in the doped Hubbard 
model, real materials are characterized by Coulomb interactions that extend well beyond a single lattice 
site. In effective low-energy descriptions of correlated materials, such as cuprates, the parameters 
of Hubbard-type models are understood as renormalized quantities rather than bare microscopic values. 
From this perspective, it is natural to investigate the influence of
extended nearest-neighbor and longer-range interactions, which magnitude depend strongly 
on the balance between screening, lattice effects, and electronic polarization~\cite{Feiner}. 
Their relevance is supported by 
investigations of other collective phenomena in strongly correlated electron systems, such as itinerant 
ferromagnetism, metal–insulator transitions, electronic ferroelectricity, and charge and spin density 
waves~\cite{Strack1,Strack2,Amadon,Farky1,Farky2}, where substantial effects on stabilization of these 
phenomena, arising from nonlocal interactions, have been reported.

Simple intuitive arguments suggest that repulsive nearest-neighbor interactions should suppress 
d-wave superconductivity, since d-wave Cooper pairs are formed predominantly between electrons 
on neighboring sites. Explicit calculations confirm that repulsive nearest-neighbor interactions 
reduce superconducting correlations; however, this suppression does not necessarily imply an 
immediate destabilization of the superconducting ground state. Although some studies argue that 
arbitrarily weak extended interactions destroy superconductivity~\cite{Alex}, a number of numerical 
and analytical works indicate that superconductivity renains stable until interactions reach some 
threshold~\cite{Gazza,Onari,Raghu2,Plakida,Senechal}. Moreover, extended interactions do not act solely 
on the pairing channel. They also strongly influence spin and charge correlations, which may either 
cooperate with superconductivity or compete against it by stabilizing alternative ordered states~\cite{Su,Zanardi}. 
Disentangling these competing tendencies is essential for assessing the robustness of superconductivity in realistic 
correlated systems and for understanding how nonlocal Coulomb interactions reshape the phase diagram of 
the Hubbard model.

In previous theoretical studies of the influence of nonlocal interactions on superconductivity in 
extended Hubbard models, a predominantly separable approach has been adopted, in which the effects 
of individual interaction terms were investigated independently. Within this framework, a wide variety 
of nonlocal interactions and processes have been examined with the aim of identifying their roles 
in stabilizing or suppressing superconductivity. Among the most frequently studied contributions are 
nearest-neighbor (i) Coulomb interaction, (ii) correlated hopping, (iii) exchange interaction, 
(iv) pair hopping, as well as next-nearest-neighbor (v) electron hopping and (vi) Coulomb interaction.
Within this separable approach, both positive~\cite{Hirsch1,Hirsch2,Hirsch3,Lara,Aligia,Bulka,Hori} and
negative~\cite{Alex,Gazza,Onari,Raghu2,Plakida,Senechal,Plonka} effects on superconductivity have been reported. 
However, the neglect of additional interaction terms, often of comparable magnitude, raises questions 
regarding the robustness and internal consistency of these conclusions. In realistic strongly correlated 
materials, multiple nonlocal interactions are simultaneously present and may interfere in a nontrivial 
manner. Treating them in isolation may therefore provide an incomplete or even misleading picture of the 
stability of the superconducting state.

In the present work, we adopt a more comprehensive approach. In addition to nearest-neighbor electron 
hopping and the on-site Coulomb interaction of the standard Hubbard model, we explicitly include 
next-nearest-neighbor electron hopping (v) together with at least one additional nonlocal interaction 
from the set (i)–(iv). The inclusion of the next-nearest-neighbor hopping term is motivated by its 
well-established relevance in high-temperature cuprate superconductors, while the remaining terms model 
the individual or combined effects of several nonlocal interactions that are likewise physically present 
in these materials. From this perspective, the approach presented in this paper represents one of the most 
comprehensive treatments of superconductivity in strongly correlated systems within the extended Hubbard 
framework. Crucially, our study is based on numerically exact calculations and, in contrast to several 
earlier works, does not focus exclusively on a single pairing symmetry. Instead, we systematically 
investigate the influence of nonlocal interactions on both s-wave and d-wave pairing channels in parallel.
The primary limitation of our approach is the relatively small system size accessible to exact 
diagonalization, namely a $L = 4 \times 4$ cluster. In order to partially mitigate this limitation, 
we supplement our exact results with additional numerical calculations using the projector quantum 
Monte Carlo (PQMC) method for selected parameter regimes of the extended Hubbard model. This combined 
strategy allows us to assess the robustness of our conclusions beyond small clusters and strengthens 
the relevance of our findings for the physics of extended, strongly correlated systems.

\section{Model and Methods}
To examine the effects of nonlocal interactions/processes on superconducting correlations 
in the extended Hubbard model we consider a tight binding Hamiltonian for a single
band. Keeping only matrix elements between nearest-neighbor Wannier states, the kinetic 
and potential energy terms result in the Hamiltonian:
\begin{align} 
H &{} 
= H_t+H_{t_n}+H_{U}+H_{V}+H_{t_c}+H_{J}+H_{J_c}= \\
& - t\sum_{\langle i,j\rangle,\sigma} c_{i\sigma}^\dagger c_{j\sigma}
- t_n\sum_{\langle \langle i,j\rangle \rangle,\sigma} c_{i\sigma}^\dagger
c_{j\sigma}
+ U \sum_i n_{i\uparrow} n_{i\downarrow}
+ V \sum_{\langle i,j\rangle} n_i n_j \nonumber \\
& + t_c \sum_{\langle i,j\rangle,\sigma}
\left( n_{i,-\sigma} + n_{j,-\sigma} \right)
c_{i\sigma}^\dagger c_{j\sigma} 
+ J \sum_{\langle i,j\rangle,\sigma,\sigma'}
c_{i\sigma}^\dagger c_{j\sigma'}^\dagger
c_{i\sigma'} c_{j\sigma} \nonumber \\
& + J_c \sum_{\langle i,j\rangle,\sigma,\sigma'}
c_{i\sigma}^\dagger c_{i\sigma'}^\dagger
c_{j\sigma'} c_{j\sigma},
\end{align}
where $c^\dagger_{i,\sigma}(c_{i,\sigma})$ creates (annihilates) an electron
of spin $\sigma$ in Wanier state $\Phi_i$. The single particle hopping is given
by~\cite{Cyrot}
\begin{equation}
t_{ij}=\int\Phi^{*}(\vec{r}-\vec{R}_i)\left[-\frac{\hbar^2}{2m}
\nabla^2+V \right]\Phi(\vec{r}-\vec{R}_j)d^3\vec{r}\; , 
\label{eq:h7}
\end{equation}
where $V$ represents the nuclear potential acting on the electrons.
The interaction, in terms of the general matrix element
\begin{eqnarray}
\langle
ij|{\frac{1}{r}}|kl\rangle &=& e^2\int\Phi^{*}(\vec{r}-\vec{R}_i)\Phi^{*}(\vec{r}\,'-\vec{R}_j)
\nonumber\\
&\times &\frac{1}{|\vec{r}-\vec{r}\,'|}\Phi(\vec{r}-\vec{R}_k)\Phi(\vec{r}-\vec{R}_l)d^3\vec{r}d^3\vec{r}\,' .
\label{eq:h8}
\end{eqnarray}
are given by: 

\begin{eqnarray}
U=\langle ii|1/r|ii\rangle &
\\
t_c=\langle ii|1/r|ij\rangle & 
\\
V=\langle ij|1/r|ij\rangle & 
\\
J=\langle ij|1/r|ji\rangle & 
\\
J_c=\langle ii|1/r|jj\rangle & 
\label{eq:h9}
\end{eqnarray}
The first two terms in Eq. (1) describe the kinetic energy of itinerant electrons 
hopping between nearest- and next-nearest-neighbor sites with amplitudes $t$ and $t_n$, 
respectively. The next terms correspond to the on-site Coulomb interaction $U$,
nearest-neighbor Coulomb interaction ($V$), correlated hopping interaction ($t_c$), 
exchange interaction ($J$) and pair hopping interaction ($J_c$). All matrix elements in
Eq.~1 are expected to be allways positive, except possibly for the matrix 
element $t_c$.   

The ground states of the model (1) are obtained using the Lanczos exact diagonalization 
method~\cite{Dag}. The numerical calculations are performed on a finite $L = 4 \times 4$ 
cluster at electron filling $N_{\uparrow} = N_{\downarrow} = 5$, corresponding to 
a hole doping of $\delta = 0.375$. To assess the robustness of our results, we complement 
the exact diagonalization data with PQMC simulations.
in selected regions of the parameter space of Hamiltonian~(1).

Having determined the ground state of Hamiltonian (1) for a given set of model 
parameters $t, t_n, U, V, J, J_c,$ and $t_c$, the superconducting correlations 
can be evaluated directly. A standard approach to study superconductivity 
in Hubbard-type models is to analyze two-particle correlation functions with 
well-defined pairing symmetries and to examine the emergence of off-diagonal 
long-range order~\cite{Yang}.
To this end, we compute the equal-time pair–pair correlation functions 
with $s$- and $d$-wave symmetry using the exact-diagonalization~\cite{Dag} and PQMC 
method~\cite{Sorella,Loh,Imada}. Our definitions follow those employed in previous 
studies~\cite{Fettes,Husslein,Fark3,Fark4,Fark5}. The corresponding correlation functions 
are given by:
\begin{equation}
C_s(r)=\frac{1}{L}\sum_{i}\langle
d^+_{i\uparrow}d^+_{i\downarrow}d_{i+r\downarrow}d_{i+r\uparrow}
\rangle,
\end{equation}
\begin{equation}
C_d(r)=\frac{1}{L}\sum_{i,\delta,\delta'}g_{\delta}g_{\delta'}\langle
d^+_{i\uparrow}d^+_{i+\delta\downarrow}d_{i+\delta'+r\downarrow}d_{i+r\uparrow}
\rangle,
\end{equation}
where $g_{\delta} = +1$ for bonds oriented along the $x$ direction 
and $g_{\delta} = -1$ for those along the $y$ direction. The sums over 
$\delta$ and $\delta'$ extend over the nearest neighbors of site $i$.

However, we emphasize that the correlation functions $C_s(r)$ and $C_d(r)$ 
defined above do not constitute ideal measures of superconducting order, 
as they generally contain contributions originating from single-particle correlations
\begin{equation}
C^{\sigma}_0(r)=\frac{1}{L}\sum_{i}\langle d^+_{i\sigma}d_{i+r\sigma}
\rangle,
\end{equation}
that provide nonzero contributions to $C_s(r)$ and $C_d(r)$ 
even in the non-interacting case. 

For this reason, the vertex correlation functions $C^{v}_s(r)$ and $C^{v}_d(r)$
in the $s$- and $d$-wave channels and their averages $C^{v}_{\alpha}$ are more commonly 
employed in the literature to identify superconducting order. 
These are defined as:
\begin{equation}
C^{v}_s(r)=C_s(r)-C^{\uparrow}_0(r)C^{\downarrow}_0(r)
\rangle,
\end{equation}
\begin{equation}
C^{v}_d(r)=C_d(r)-\sum_{\delta,\delta'}g_{\delta}g_{\delta'}
C^{\uparrow}_0(r)C^{\downarrow}_0(r+\delta-\delta')
\rangle,
\end{equation}

\begin{equation}
C^{v}_{\alpha}=\frac{1}{L}\sum_{i}C^{v}_{\alpha}(i),
\end{equation}
where $\alpha$=s,d.
It should be noted that while in the thermodynamic limit $L \to \infty$ non-zero values 
of correlation functions indicate the occurrence of superconducting state in the system, 
on finite clusters they can be used only as precursors of superconductivity.

\section{Results and discussion}
\subsection{Effects of $U$}
To reveal the impact of the interaction terms introduced above on the
superconducting correlation functions in the extended Hubbard model, we begin 
by considering effects of the on-site Hubbard interaction ($H=H_t+H_{t_n}+H_U$). 
Figure~1 presents the results of our exact
diagonalization calculations for the $s$-wave and $d$-wave superconducting
correlation functions, $C_s$ and $C_d$, shown as $t_n$-$U$ phase diagrams
obtained on the $L=4\times4$ cluster at electron filling
$N_\uparrow=N_\downarrow=5$.
\begin{figure}[h!]
\begin{center}
\includegraphics[width=7cm]{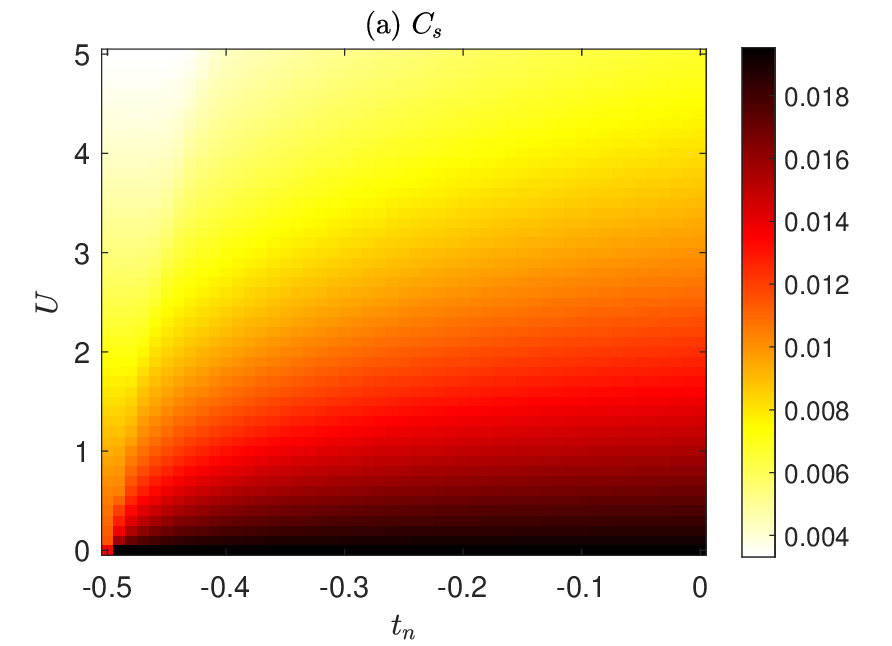}
\includegraphics[width=7cm]{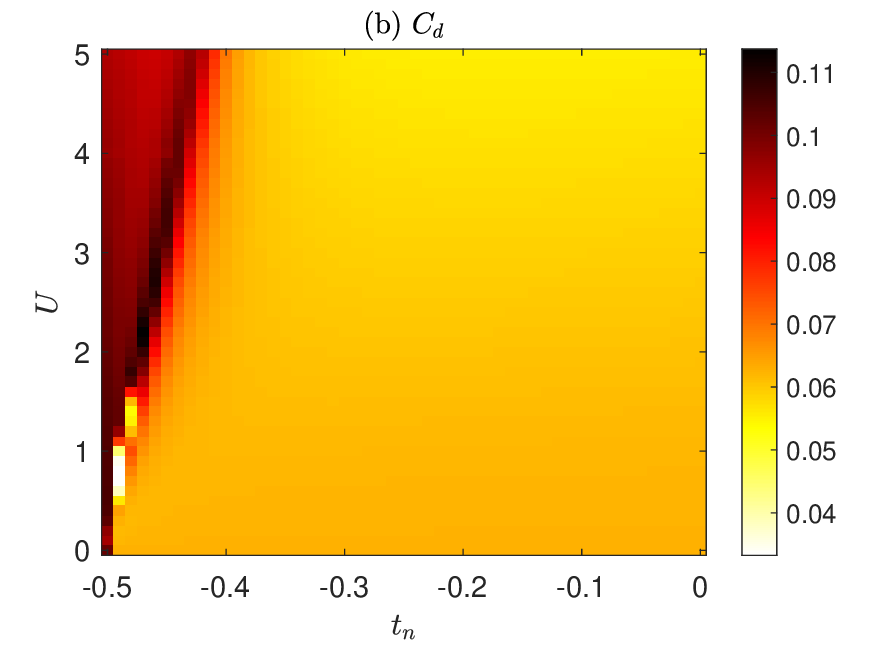}
\includegraphics[width=7cm]{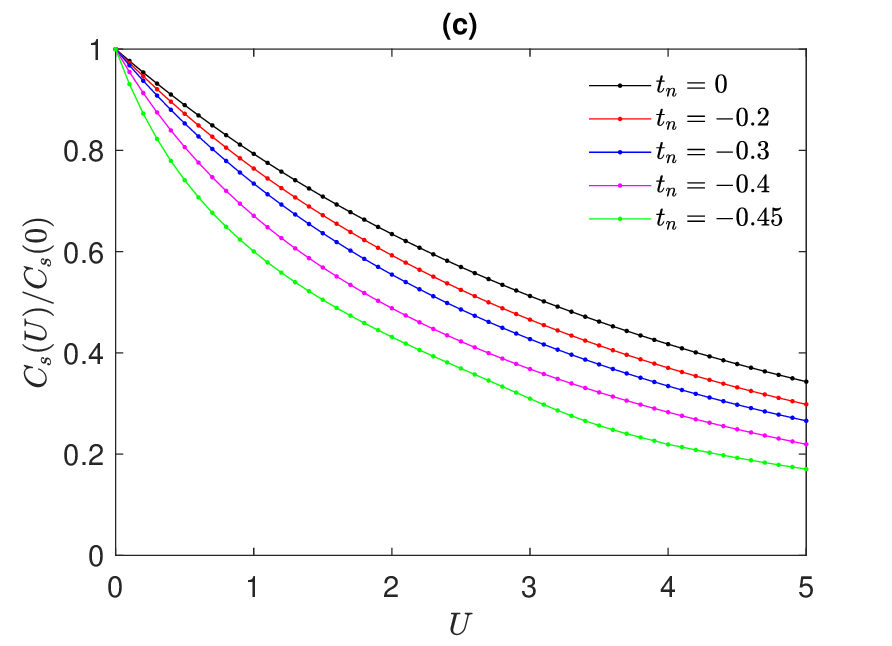}
\includegraphics[width=7cm]{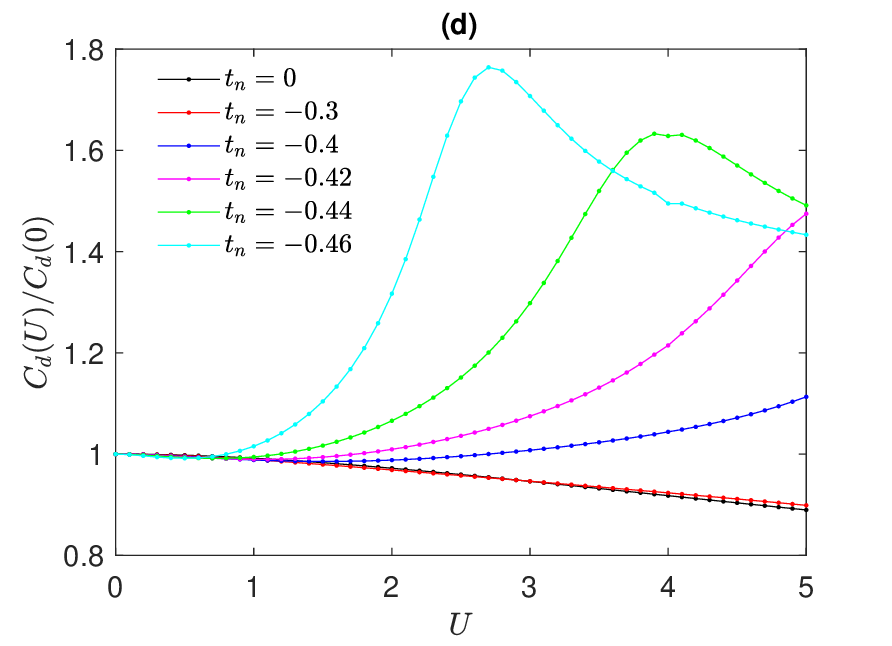}
\end{center}
\caption{\small Correlation functions $C_s$ (a) and $C_d$ (b) with s- and
d-wave symmetry as functions of $U$ and $t_n$  calculated for
the $L=4 \times 4$ cluster and $N_{\uparrow}=N_{\downarrow}=5$. The ratio $C_s(U)/C_s(U=0)$ (c) 
and  $C_d(U)/C_d(U=0)$ (d) as a function of $U$ calculated for different 
values of $t_n$.}
\label{fig1}
\end{figure}
Several general trends can be identified. The on-site Hubbard interaction
strongly suppresses superconducting correlations in the $s$-wave channel
(see Fig.~1(a)) and this suppression is further enhanced with increasing
$|t_n|$, as illustrated in Fig.~1(c). In particular, the effect of $U$ is
substantial and leads to a pronounced reduction of $C_s(U)$ by approximately
a factor of five relative to its noninteracting value $C_s(U=0)$.
In contrast, a qualitatively different behavior is observed for the
$d$-wave superconducting correlation function $C_d$, shown in
Figs.~1(b) and 1(d). For small values of $|t_n|$ ($|t_n|<0.3$), $C_d$ is
gradually suppressed with increasing $U$, similarly to the $s$-wave case.
However, for $|t_n|>0.3$, $C_d$ increases sharply, reaches a maximum, and
subsequently decreases again at larger $|t_n|$. The maximal enhancement of
$C_d(U)$ with respect to its noninteracting value $C_d(U=0)$ is approximately
a factor of two, indicating that the on-site Hubbard interaction plays an
important role in stabilizing $d$-wave superconducting correlations.
It should be emphasized, however, that this enhancement of $C_d(U)$ is not a
pure effect of the Hubbard interaction alone, but rather arises from a
cooperative interplay between $U$ and the next-nearest-neighbor hopping
amplitude $t_n$.

\subsection{Effects of $t_c$}
Let us now investigate the effects of the so-called correlated hopping term
$t_c$ ($H=H_t+H_{t_n}+H_U+H_{t_c}$).
The results of our exact diagonalization calculations obtaimed for $U=4$ and
$L=4\times 4$ are presented in Fig.~2. 
At first glance, it is evident that its impact on the superconducting
correlation functions in the $s$- and $d$-wave channels is markedly different
from the previously discussed case of repulsive on-site interaction $U>0$.
\begin{figure}[h!]
\begin{center}
\includegraphics[width=7cm]{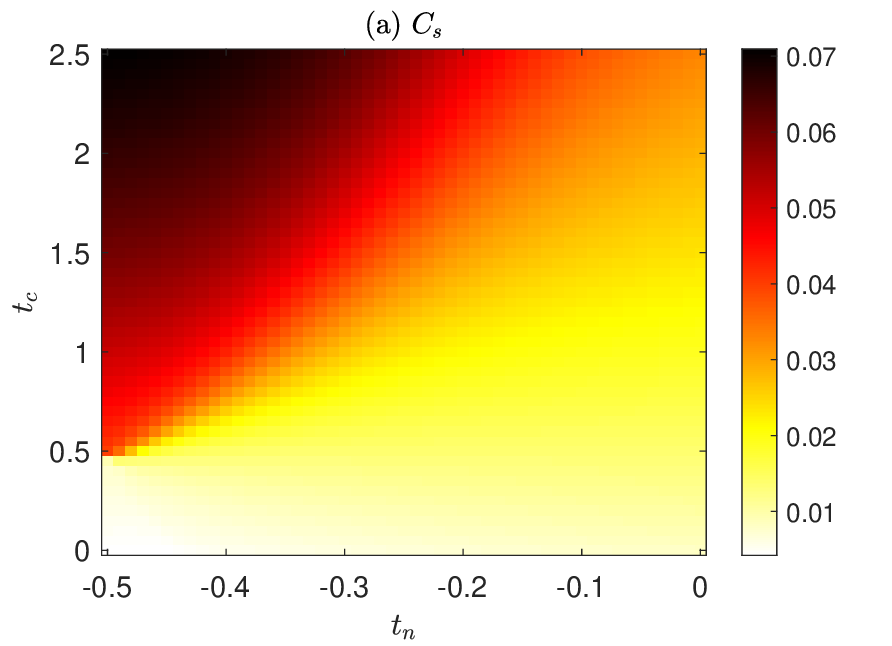}
\includegraphics[width=7cm]{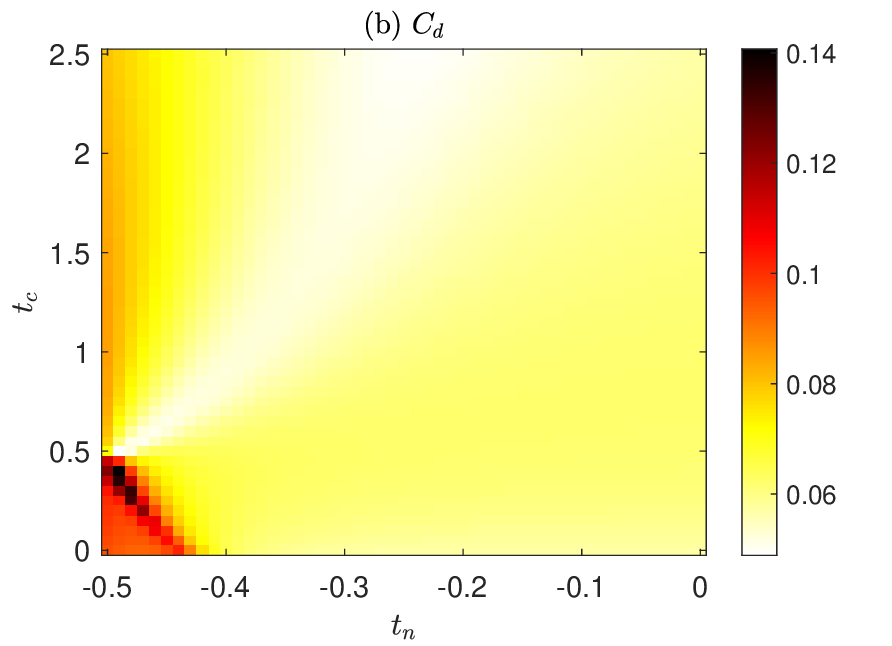}
\includegraphics[width=7cm]{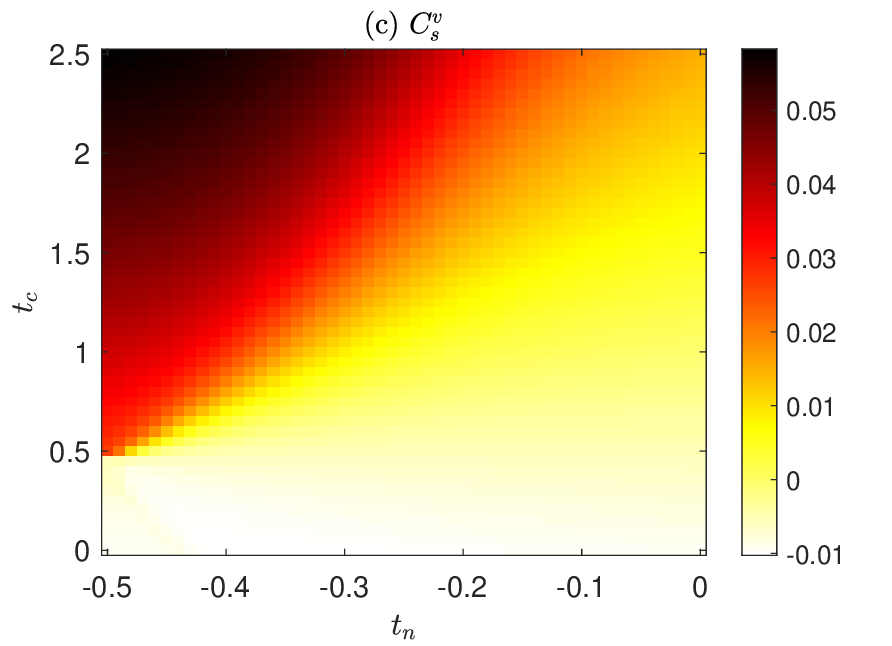}
\includegraphics[width=7cm]{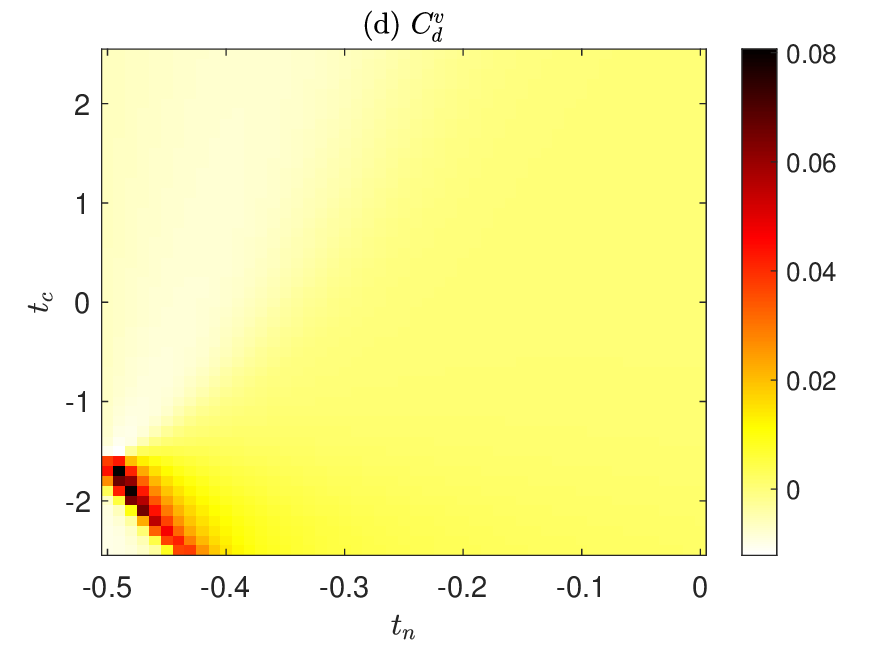}
\includegraphics[width=7cm]{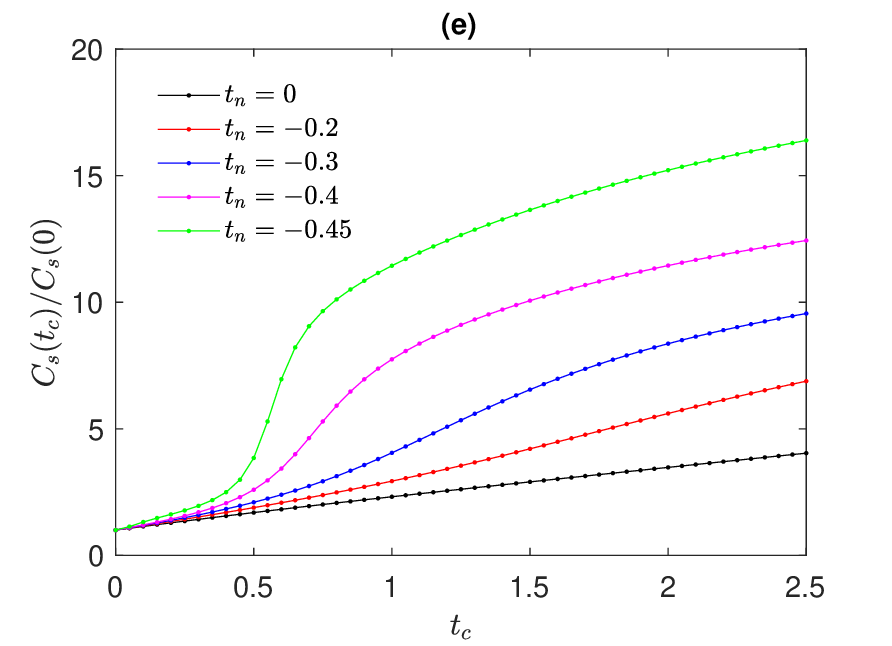}
\includegraphics[width=7cm]{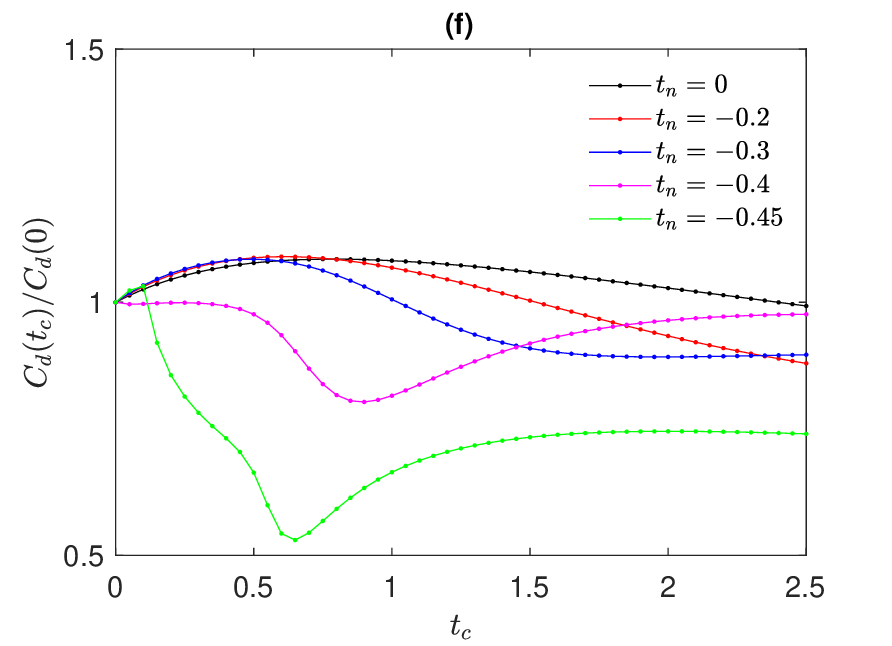}
\end{center}
\caption{\small Correlation functions $C_s$ (a) and $C_d$ (b) with s- and
d-wave symmetry as functions of $t_c$ and $t_n$  calculated for $U=4$,
$N_{\uparrow}=N_{\downarrow}=5$ on the $L=4 \times 4$ cluster. 
The vertex correlation functions $C^v_s$ (c) and $C^v_d$ (d) 
as functions of $t_c$ and $t_n$.
The ratio $C_s(t_c)/C_s(t_c=0)$ (e) 
and  $C_d(t_c)/C_d(t_c=0)$ (f) as a function of $t_c$ calculated for different 
values of $t_n$.}
\label{fig2}
\end{figure}
A~significant enhancement of superconducting correlations is now observed in
the $s$-wave channel and is again strongly supported by increasing $|t_n|$.
Very large values of $C_s$ induced by the correlated hopping term
$t_c$ appear at the boundary of the parameter range considered here
($t_n=-0.5$), where the ratio $C_s(t_c)/C_s(t_c=0)$ reaches values of the order of $15$. 
This demonstrates that correlated hopping term constitutes a key ingredient for the 
stabilization of superconducivity in strongly correlated systems and should
be taken into account in the correct description of these systems.
In contrast to the $s$-wave correlation function $C_s$, the $d$-wave
correlation function $C_d$ exhibits only very weak changes as the correlated
hopping amplitude $t_c$ is increased. Depending on the value of $t_n$, $t_c$
either slightly enhances or suppresses superconducting correlations in the
$d$-wave channel. Specifically, a moderate enhancement is observed for small
values of $|t_n|$ ($|t_n|<0.4$), while a suppression occurs in the opposite
limit. 

In addition to the correlation functions $C_s$ and $C_d$, Fig.~2 also 
shows the vertex correlation functions $C_s^v$ and $C_d^v$ (Figs.~2(c)
and 2(d)). A direct comparison of the phase diagrams corresponding to
$C_s$ and $C_s^v$, as well as to $C_d$ and $C_d^v$, reveals that they are
qualitatively identical. Since the same observation holds for all remaining
interaction terms introduced in the previous section, in the following we
present only the results for the bare correlation functions $C_s$ and $C_d$.
Negative values of $C_s$ and $C_d$ indicate persistent lattice effects, which
render quantitative comparison ambiguous.

In accordance with several previous studies that have examined the effect 
of negative values of the parameter $t_c$ on the ground-state properties 
of the extended Hubbard model, we have also carried out the same numerical 
analysis of Hamiltonian (1) for $t_c < 0$. Our results indicate that 
superconducting correlations in both the s-wave and d-wave channels are 
significantly reduced in this regime, by approximately a factor of three 
to four, compared to the case of positive $t_c$. In view of this pronounced 
suppression, the $t_c < 0$ regime will not be considered further in the present 
study.
\subsection{Effects of $V$}
The effects of the nearest-neighbor Coulomb interaction $V$
($H=H_t+H_{t_n}+H_U+H_{V}$) on the superconducting correlation 
functions $C_s$ and $C_d$ are shown in Fig.~3.
\begin{figure}[h!]
\begin{center}
\includegraphics[width=7cm]{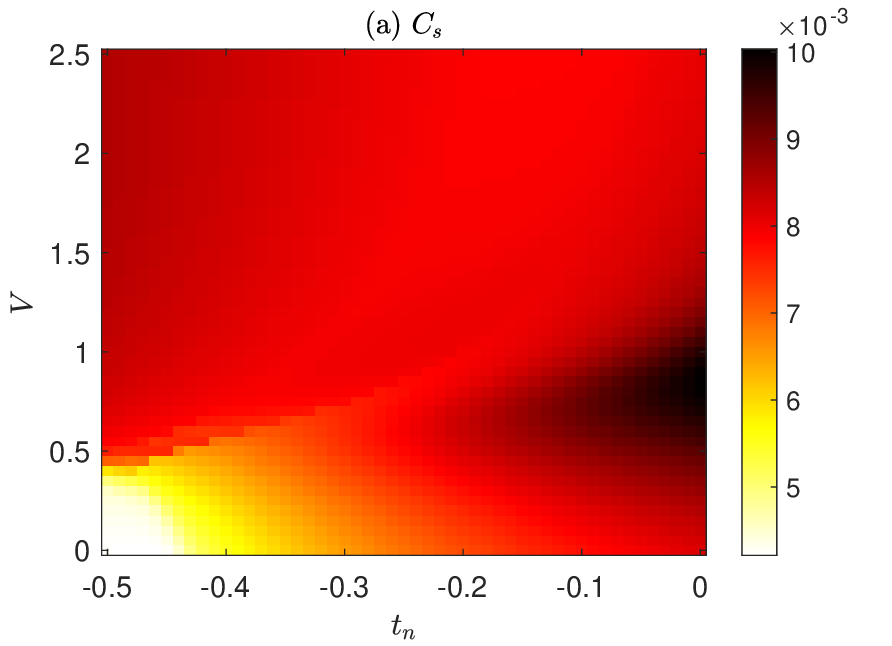}
\includegraphics[width=7cm]{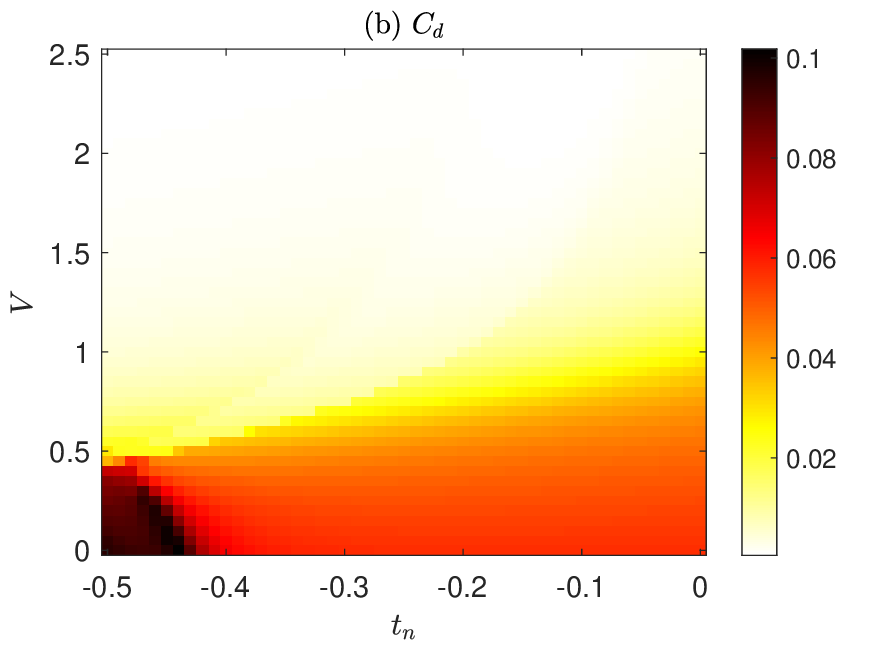}
\includegraphics[width=7cm]{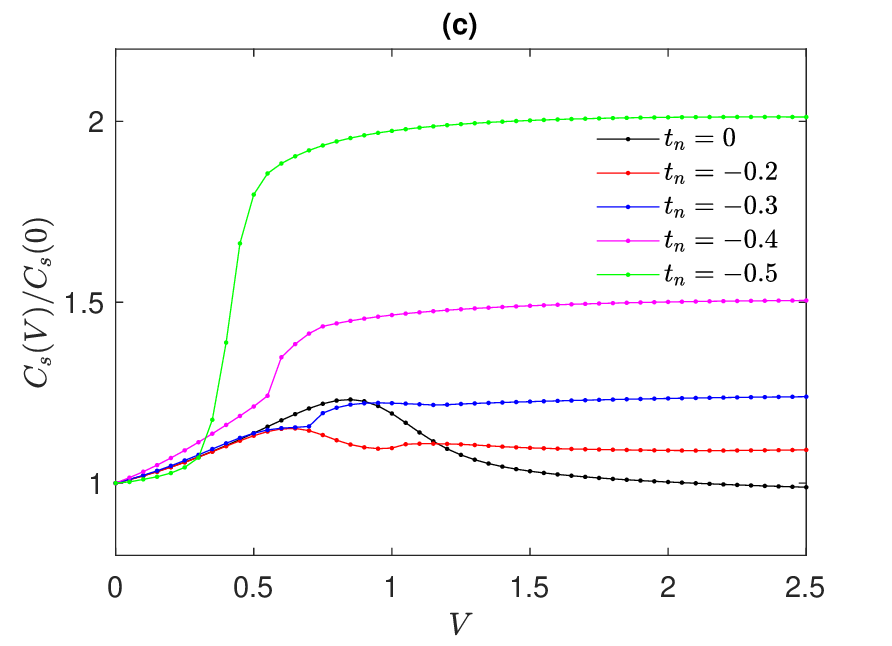}
\includegraphics[width=7cm]{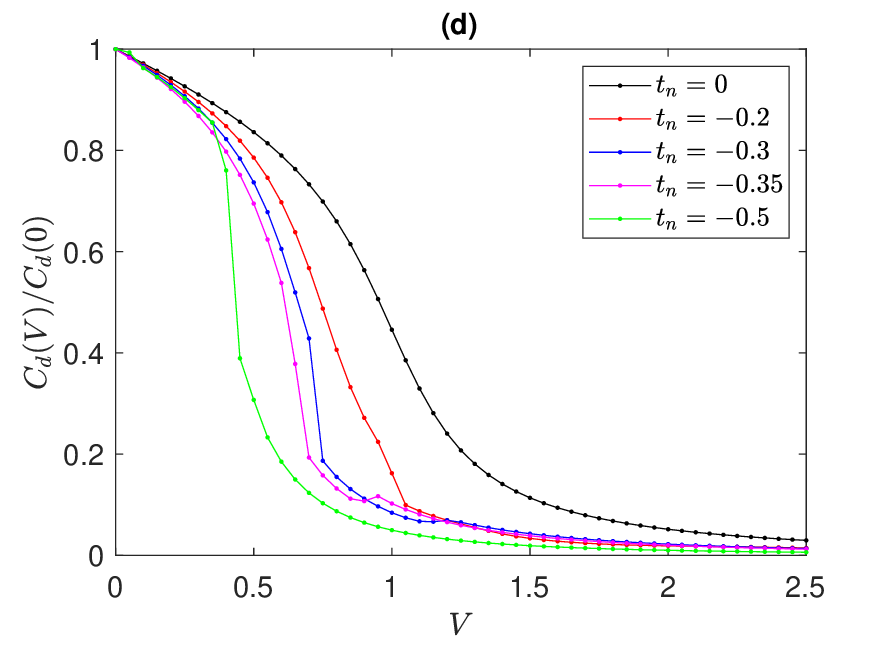}
\end{center}
\caption{\small Correlation functions $C_s$ (a) and $C_d$ (b) with s- and
d-wave symmetry as functions of $V$ and $t_n$  calculated for
$U=4$, $N_{\uparrow}=N_{\downarrow}=5$ on the $L=4 \times 4$ cluster. The ratio $C_s(V)/C_s(V=0)$ (c) 
and  $C_d(V)/C_d(V=0)$ (d) as a function of $V$ calculated for different 
values of $t_n$.}
\label{fig3}
\end{figure}
A direct comparison of the corresponding phase diagrams reveals that $V$
has opposite effects on $C_s$ and $C_d$. In particular, $C_d$ attains its maximum 
values in the vicinity of $V=0$ and $t_n=-0.5$ and is suppressed in the remaining 
regions of the phase diagram, whereas $C_s$ reaches its minimum near this point 
and is enhanced elsewhere.
Cuts through the $C_s$ phase diagram at fixed $t_n$ [Fig.~3(c)] show that
the ratio $C_s(V)/C_s(V=0)$ reaches its largest values, of the order of $2$,
for $|t_n|=0.5$. On the other hand, cuts through the $C_d$ phase diagram at fixed $t_n$
[Fig.~3(d)] show a rapid decrease of the superconducting correlations from
their initial value at $V=0$ down to zero.
\subsection{Effects of $J$}
The effects of the exchange interaction $J$ ($H=H_t+H_{t_n}+H_U+H_{J}$) on the
superconducting correlation functions $C_s$ and $C_d$ are shown in Fig.~4.
\begin{figure}[h!]
\begin{center}
\includegraphics[width=7cm]{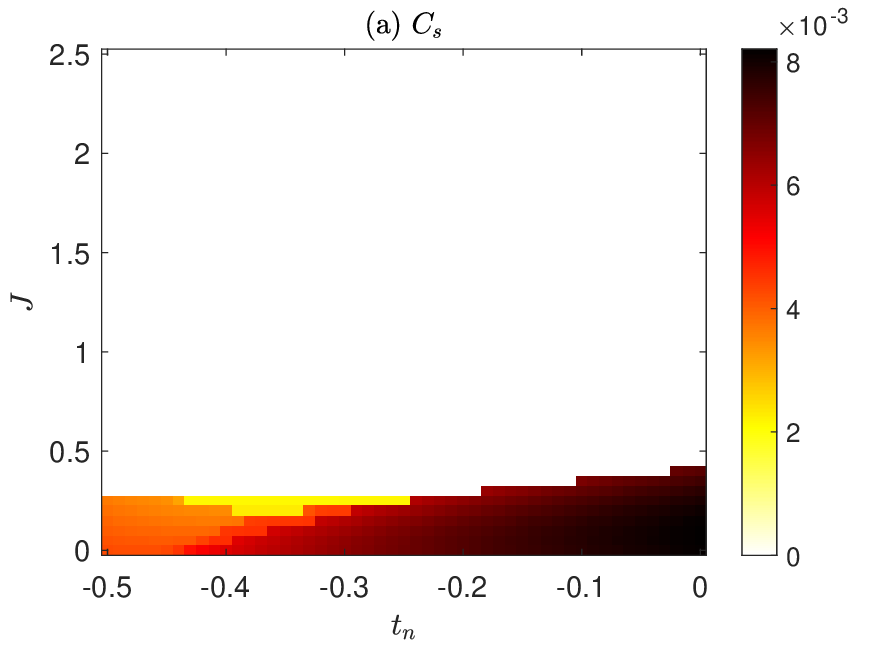}
\includegraphics[width=7cm]{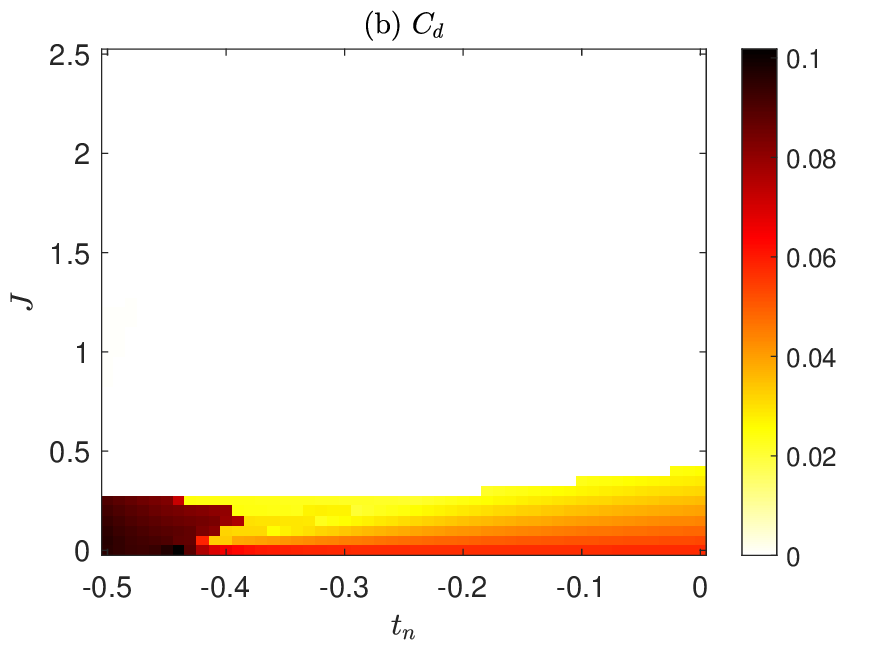}
\includegraphics[width=7cm]{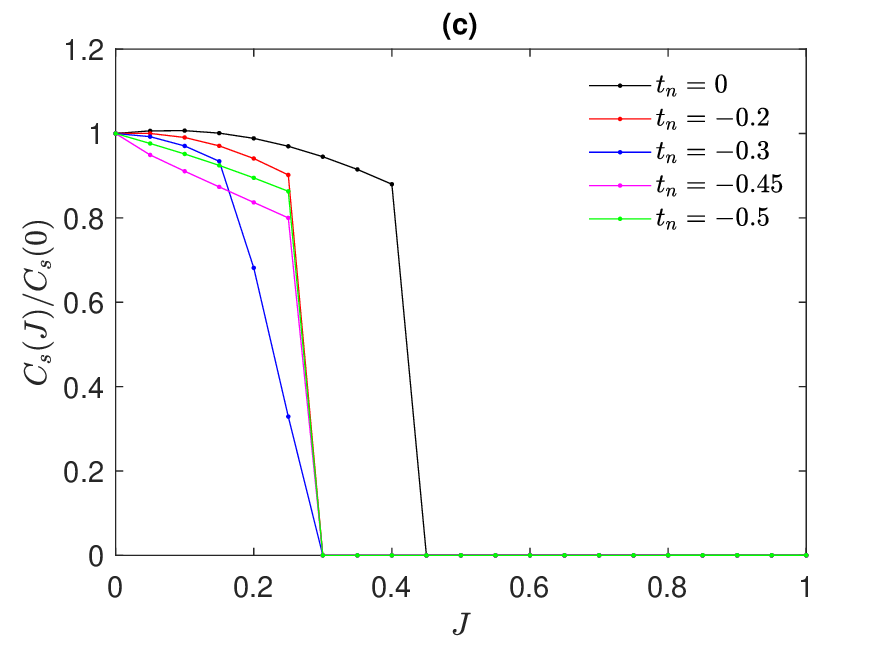}
\includegraphics[width=7cm]{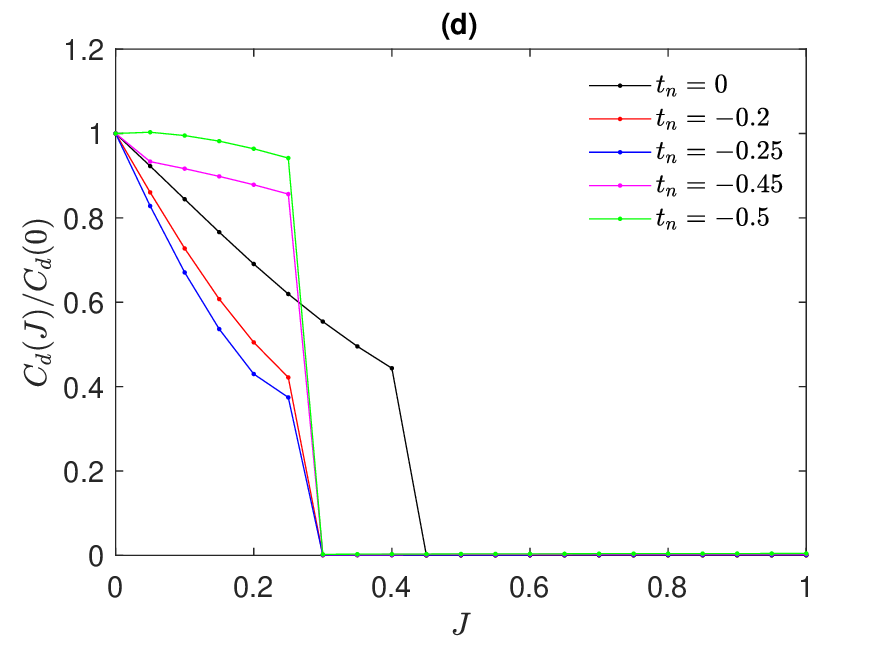}
\end{center}
\caption{\small Correlation functions $C_s$ (a) and $C_d$ (b) with s- and
d-wave symmetry as functions of $J$ and $t_n$  calculated for
$U=4$, $N_{\uparrow}=N_{\downarrow}=5$ on the $L=4 \times 4$ cluster. The ratio $C_s(J)/C_s(J=0)$ (c) 
and  $C_d(J)/C_d(J=0)$ (d) as a function of $J$ calculated for different 
values of $t_n$.}
\label{fig4}
\end{figure}
We find that the exchange interaction $J$ produces qualitatively similar
effects on the phase diagrams of $C_s$ and $C_d$. Both $C_s$ and $C_d$ are
finite at $J=0$, decrease gradually with increasing $J$, and then vanish
discontinuously at a critical value $J_0$. The critical values $J_0$ are
relatively small, approximately an order of magnitude smaller than the
on-site interaction strength $U$. Nevertheless, they are sufficient to
completely suppress superconducting correlations.
This observation indicates that, although the exchange interaction $J$
appears lower in the hierarchy of interaction terms discussed in the
previous section, following $U$, $V$, and $t_c$, its role may be crucial for
a proper description of superconductivity in strongly correlated systems.
Another noteworthy result that can be inferred from the presented phase
diagrams is that $C_s$ and $C_d$ reache the largest values in two opposite
parameter regimes: $C_s$ in the limit $J \to 0$ and $t_n \to 0$, and $C_d$ in
the limit $J \to 0$ and $t_n \to -0.5$.
\subsection{Effects of $J_c$}
A qualitatively similar effect to that produced by $J$ in the previous case
is observed for the pair-hopping interaction $J_c$ ($H=H_t+H_{t_n}+H_U+H_{J_c}$)
in the $s$-wave superconducting correlation function (see Fig.~5). 
\begin{figure}[h!]
\begin{center}
\includegraphics[width=7cm]{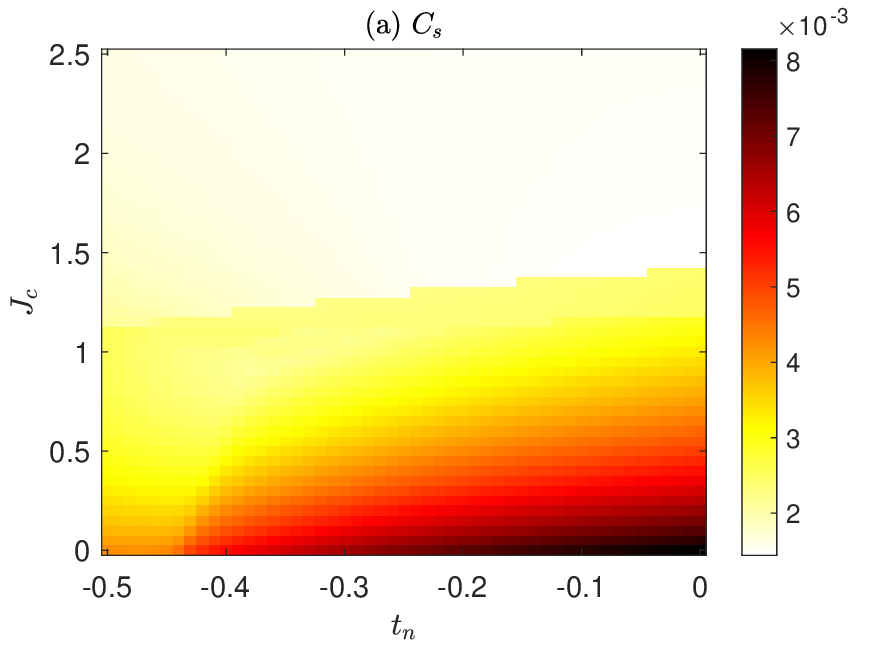}
\includegraphics[width=7cm]{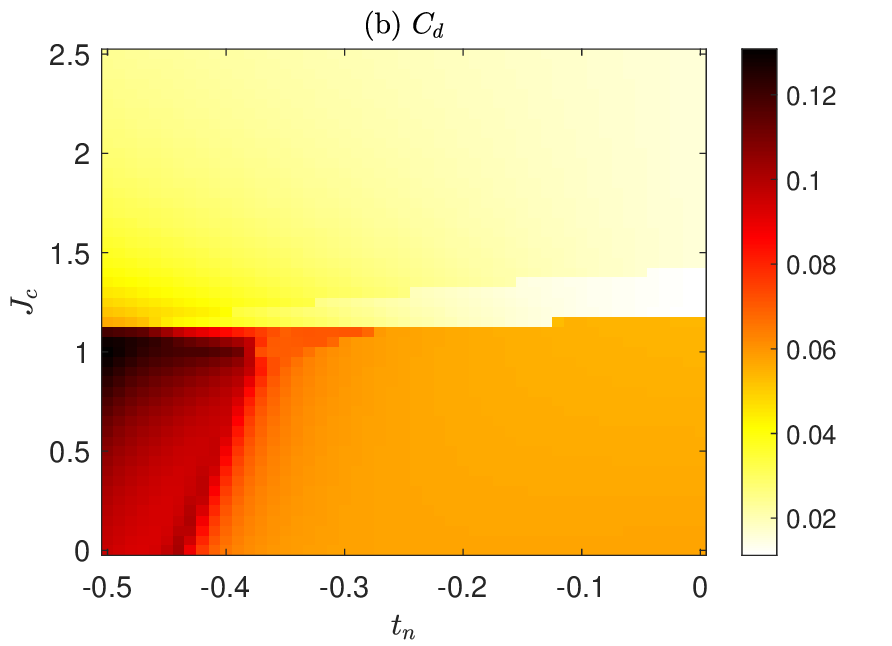}
\includegraphics[width=7cm]{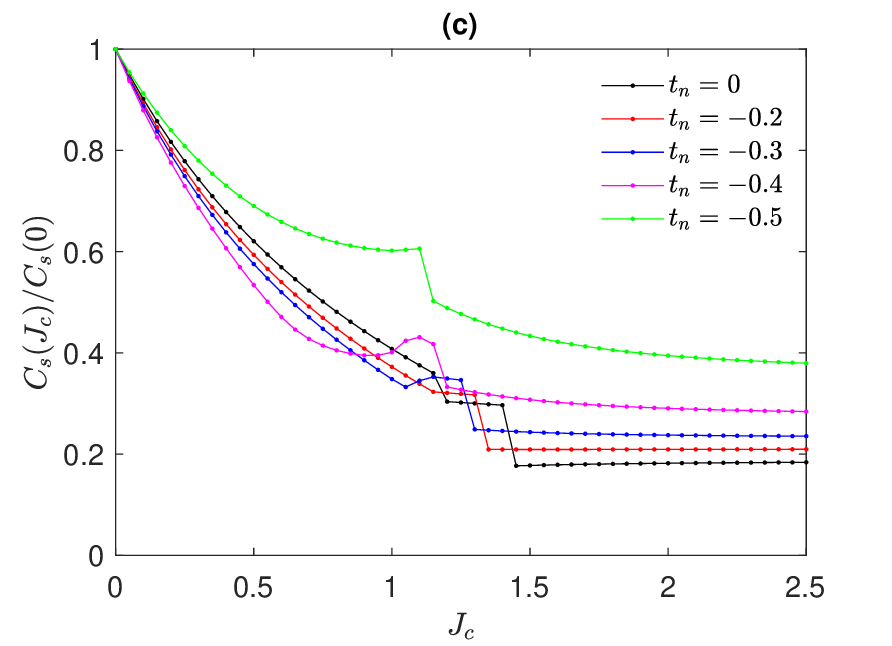}
\includegraphics[width=7cm]{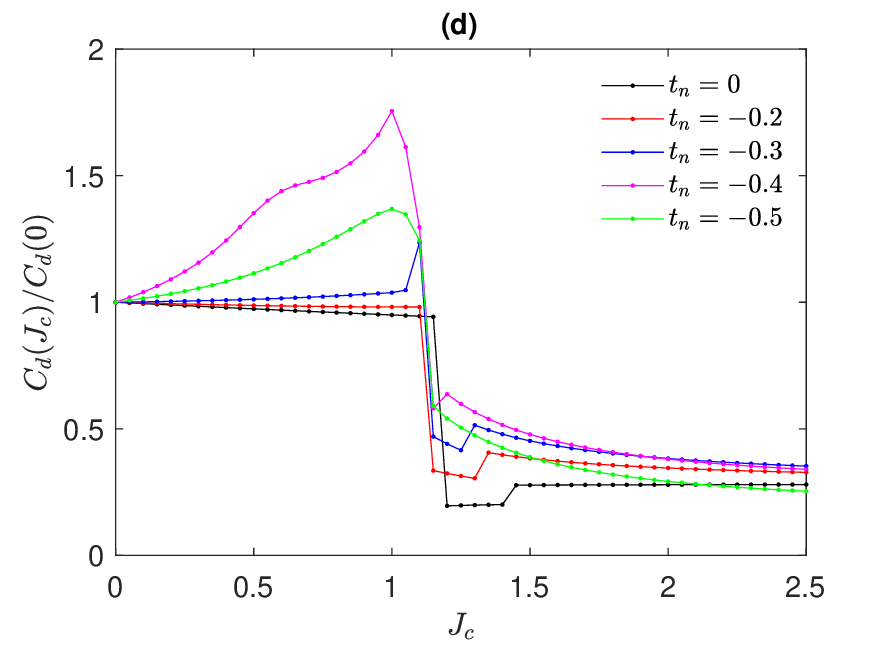}
\end{center}
\caption{\small Correlation functions $C_s$ (a) and $C_d$ (b) with s- and
d-wave symmetry as functions of $J_c$ and $t_n$  calculated for
$U=4$, $N_{\uparrow}=N_{\downarrow}=5$  on the $L=4 \times 4$ cluster. The ratio $C_s(J_c)/C_s(J_c=0)$ (c) 
and  $C_d(J_c)/C_d(J_c=0)$ (d) as a function of $J_c$ calculated for different 
values of $t_n$.}
\label{fig5}
\end{figure}
Specifically, $C_s$ initially decreases gradually with
increasing $J_c$, in close analogy to the cases discussed above. At a
critical value of $J_c$, $C_s$ exhibits a discontinuous jump; however, in
contrast to the behavior found for $J$, this jump is not accompanied by a
complete suppression of superconducting correlations. Instead, $C_s$
remains finite over a broad range of $J_c$ values.
In contrast, the $d$-wave correlation function $C_d$ varies only weakly with
increasing $J_c$ for small values of $|t_n|$ ($|t_n|<0.2$), but changes very
rapidly in the opposite limit $|t_n|>0.2$. In both cases, $C_d$ exhibits a
discontinuous jump at a critical value $J_c \sim 1$ and subsequently evolves
only gradually with further increasing $J_c$.
In this sense, the effects of $J$ and $J_c$ are very similar, with the only
notable difference being that the critical values of $J_c$ are approximately
a factor of two larger than those found for $J$.
\subsection{Combined effects of $V,t_c,J,J_c$}
So far, we have investigated only the effects of individual nonlocal interactions 
on superconducting correlations within the extended Hubbard model with finite 
nearest-neighbor hopping $t$, next-nearest-neighbor hopping $t_n$, and a finite 
on-site Coulomb interaction $U$.
To reveal the combined effect of two or more nonlocal interactions, which are 
present in real materials and are typically of comparable magnitude, we next 
examine two distinct cases: (i) $J = J_c$ and (ii) $V = t_c = J = J_c = I$.
The first condition is exactly satisfied in the case where the wave functions 
are real, which we assume throughout the paper. In contrast, the second condition 
is only approximately satisfied in real materials, since the parameters 
$V$ and $t_c$ are typically larger than $J$ and $J_c$.
However, since our primary 
interest lies in the qualitative effects of nonlocal interactions, 
this simplifying assumption is reasonable. 

The resulting phase diagrams for the $s$- and $d$-wave correlation functions 
corresponding to these two distinct cases are shown in Fig.~6. Comparing the phase 
diagrams for $C_s$ and $C_d$ obtained for $J = J_c$ (Fig~6a and Fig.~6b) with 
those corresponding to the case of an individual $J$ (Fig.~4), we observe that they 
are practically identical. This implies that the relevant physical processes affecting 
the superconducting correlations in both the $s$- and $d$-wave channels are driven 
exclusively by the exchange interaction $J$.
\begin{figure}[h!]
\begin{center}
\includegraphics[width=7cm]{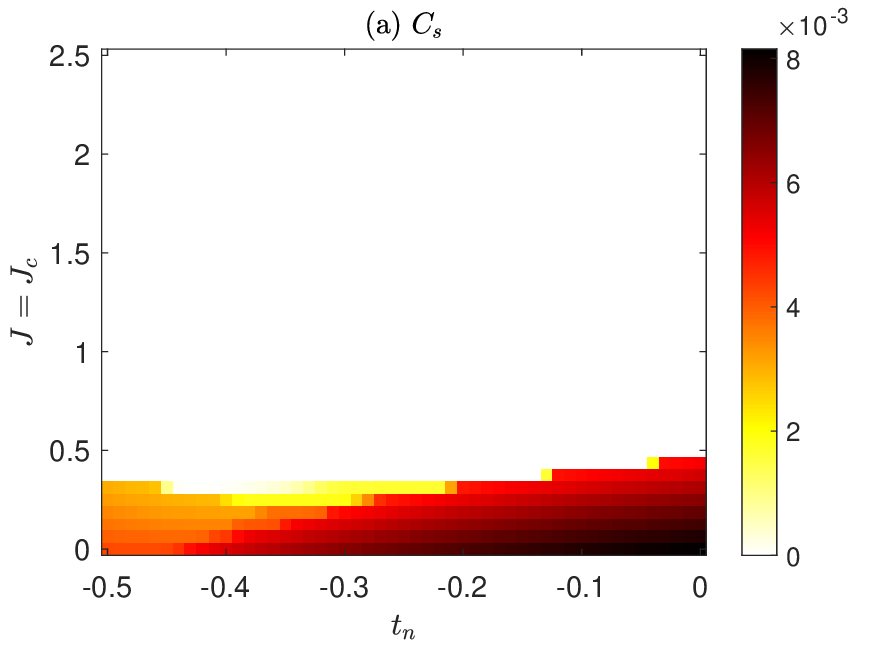}
\includegraphics[width=7cm]{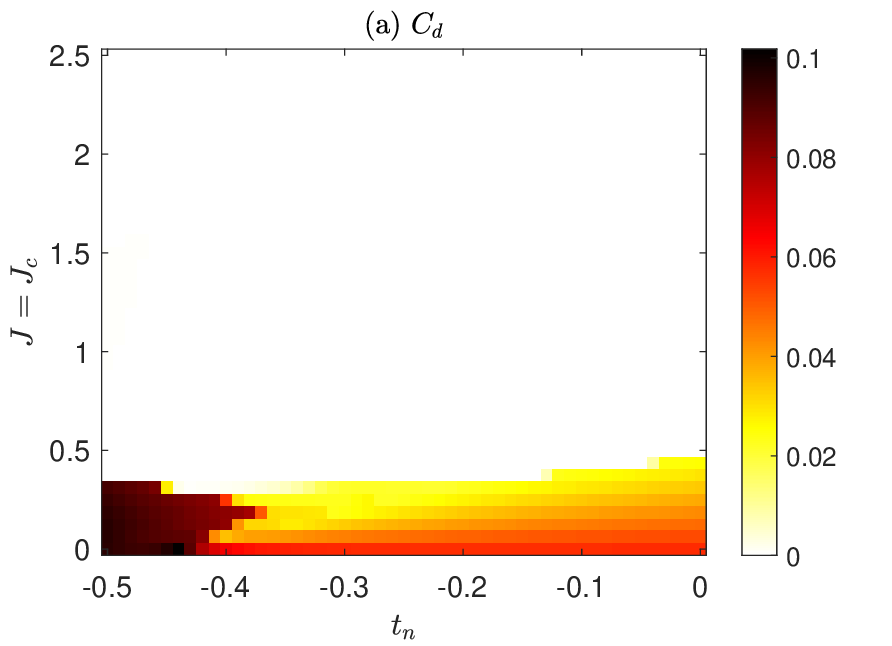}
\includegraphics[width=7cm]{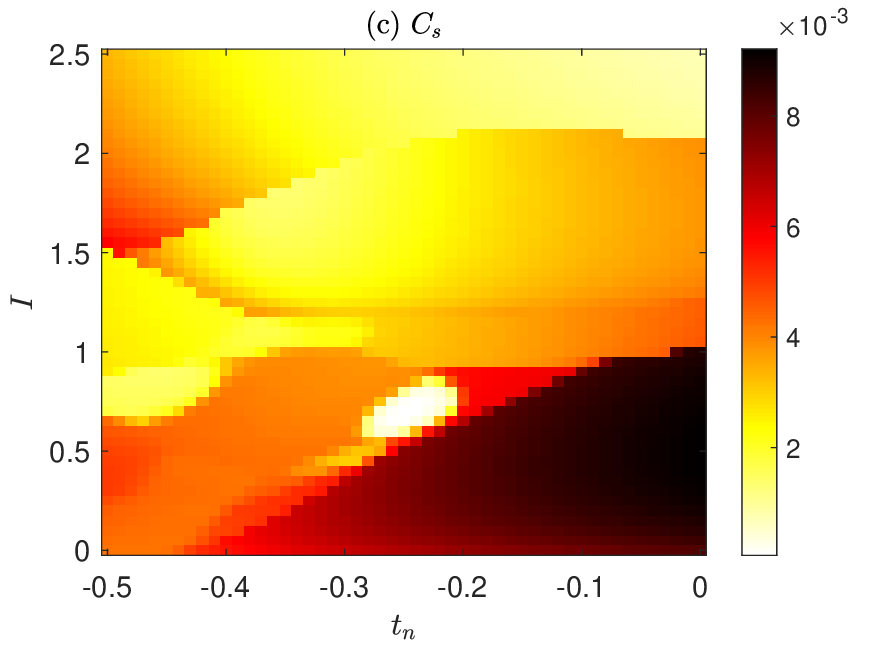}
\includegraphics[width=7cm]{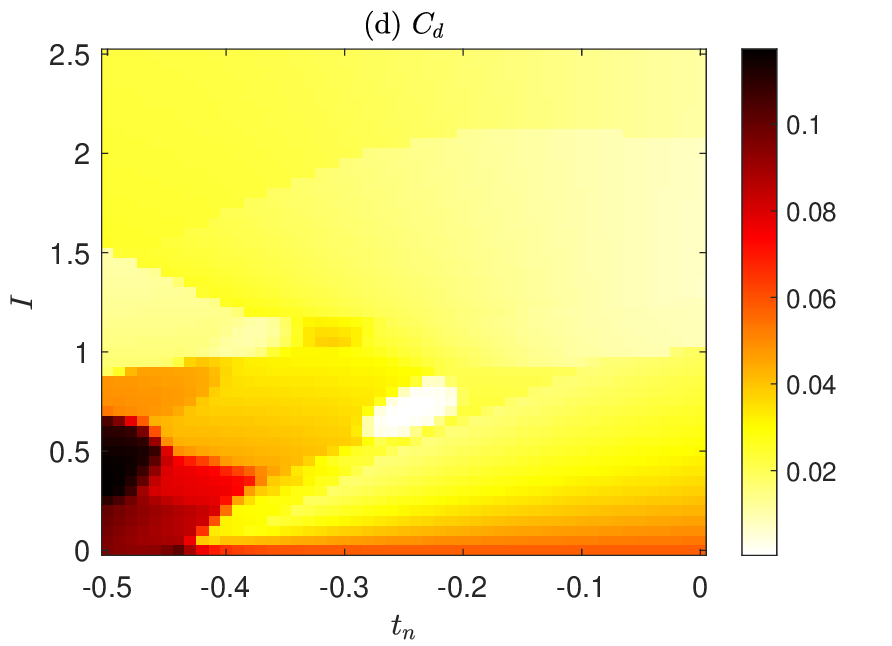}
\end{center}
\caption{\small Correlation function $C_s$  with s-wave symmetry as a function 
of $J=J_c$ and $t_n$ (a);  $I$ and $t_n$  (c) calculated for $U=4$,
$N_{\uparrow}=N_{\downarrow}=5$ on the $L=4 \times 4$ cluster. 
Correlation function $C_d$  with d-wave symmetry as a function 
of $J=J_c$ and $t_n$ (b);  $I$ and $t_n$  (d) calculated for $U=4$,
$N_{\uparrow}=N_{\downarrow}=5$ on the $L=4 \times 4$ cluster. 
}
\label{fig6}
\end{figure}
Unlike this case, the combined effects of all interactions incorporated 
in the parameter $I$ dramatically modify the picture of superconducting correlations 
in the $s$- and $d$-wave channels discussed above for individual nonlocal interactions.
The resluting phase diagrams (see Fig.~6c and Fig.~6d) demonstrate 
that, as a consequence of the combined effects of all nonlocal interactions, 
the superconducting correlations in the $s$-wave channel are strongest 
in the $t_n=0$, $I=0$ corner of the phase diagram, whereas in the $d$-wave channel 
they are strongest in the $t_n=-0.5$, $I=0$ corner of the phase diagram.
The remaining regions of the $C_s$ and $C_d$ phase diagrams display a highly 
intricate structure, consisting of numerous macroscopic domains with $C_s>0$ 
and $C_d>0$, which directly reflect the strong competition and nontrivial interplay 
among the different nonlocal interactions.

\subsection{Finite-size effects}
Since all the results discussed above were obtained on a relatively 
small cluster of $L=4\times4$ sites, it is natural to ask whether they persist 
for larger clusters as well. To address this question, at least partially, 
we have performed extensive projector quantum Monte Carlo simulations of the model 
with the Hubbard interaction term $U$ ($H=H_t+H_{t_n}+H_U$)
on the $L=12\times12$ cluster. The results of 
our numerical calculations are summarized in Fig.~7 in the form of $t_n$--$U$ phase 
diagrams for the $C_s$ and $C_d$ correlation functions. 
\begin{figure}[h!]
\begin{center}
\includegraphics[width=7cm]{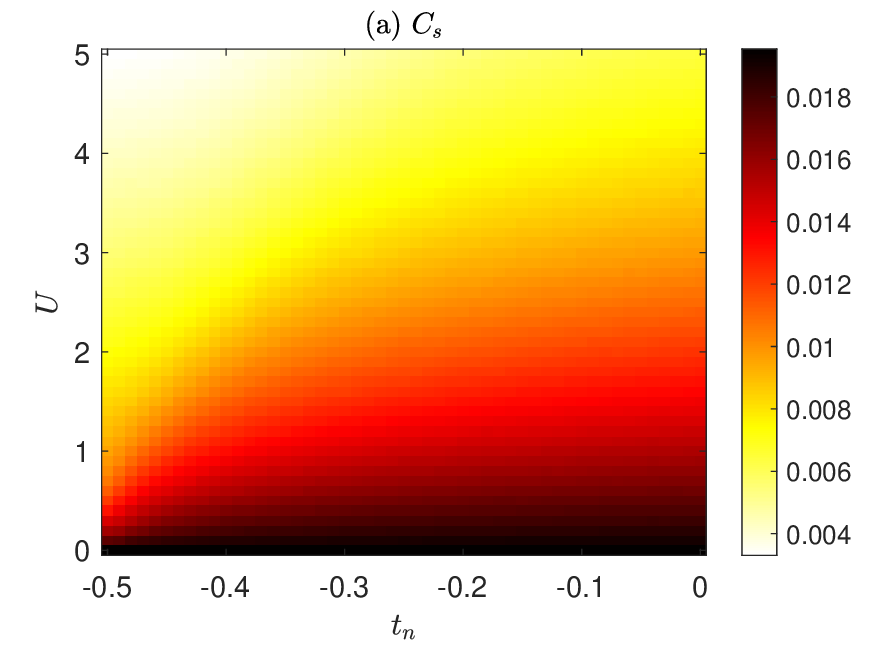}
\includegraphics[width=7cm]{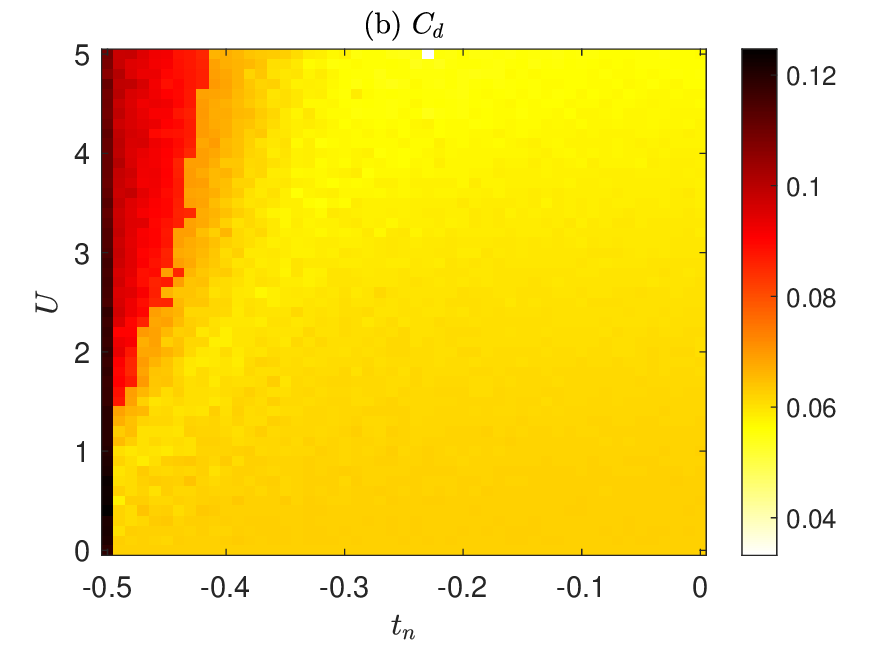}
\end{center}
\caption{\small Correlation functions $C_s$ (a) and $C_d$ (b) with s- and
d-wave symmetry as functions of $U$ and $t_n$  calculated for
$n_{\uparrow}=n_{\downarrow}=0.3125$ on the $L=12 \times 12$ cluster 
by the PQMC method.}
\label{fig7}
\end{figure}
A direct comparison of these 
phase diagrams with their exact diagonalization counterparts obtained for $L=4\times4$ 
(see Fig.~1) shows that the PQMC and exact diagonalization phase diagrams are qualitatively identical. 
This indicates that the results obtained on small clusters can be reliably extrapolated 
to larger systems, at least at a qualitative level.

\section{Conclusions}

In this work we have systematically investigated the effects of nonlocal
interactions on superconducting correlations within the extended
Hubbard model. In contrast to the predominantly separable approaches adopted
in earlier studies, where individual nonlocal terms were analyzed in isolation,
we have considered a more comprehensive framework including next-nearest-neighbor
hopping together with several physically relevant nonlocal interaction terms.
This strategy allows us to assess their mutual interplay and the resulting impact
on pairing tendencies in both the $s$- and $d$-wave channels.

Our exact diagonalization results on the $L=4\times4$ cluster, supported by
projector quantum Monte Carlo simulations for selected parameter regimes,
demonstrate that nonlocal interactions exert a highly nontrivial and symmetry-dependent
influence on superconducting correlations. While the on-site Hubbard interaction
$U$ favors $d$-wave correlations in cooperation with finite next-nearest-neighbor
hopping $t_n$, other nonlocal terms act either as strong enhancers or efficient
suppressors of pairing, depending on their character and magnitude. In particular,
the correlated hopping term $t_c$ substantially enhances $s$-wave correlations,
whereas the exchange interaction $J$ and pair-hopping term $J_c$ can drive
a rapid suppression of superconducting correlations beyond relatively small
critical values. The nearest-neighbor Coulomb interaction $V$ affects the
$s$- and $d$-wave channels in opposite ways, highlighting the intrinsic
competition between different pairing symmetries.

When all nonlocal interactions are considered simultaneously, the resulting
phase diagrams reveal a complex structure reflecting strong competition
and cooperative effects among the various interaction channels. The dominant
pairing symmetry is controlled not by a single parameter but by the collective
balance between kinetic processes and nonlocal Coulomb and exchange interactions.
This underscores the importance of treating extended interactions on equal footing
when addressing superconductivity in strongly correlated systems.

Although our exact calculations are restricted to relatively small clusters,
the qualitative agreement with PQMC results obtained on larger lattices
indicates that the main trends identified here are robust. Our findings
demonstrate that nonlocal interactions can qualitatively reshape the pairing
landscape of the extended Hubbard model and may even induce transitions
between competing pairing symmetries under suitable parameter conditions.
From a broader perspective, this study highlights that realistic modeling
of unconventional superconductors requires going beyond the minimal Hubbard
framework and incorporating the full spectrum of relevant nonlocal processes.
\vspace{1.0cm}  
\\ 
{\bf Data Availability Statement}\\ 
Data will be made available on reasonable request.
\vspace{0.5cm}
\\
{\bf Acknowledgments}\\
This work was supported by the Slovak Research and Development
Agency under the contract no. APVV-23-0226, the Slovak Grant Agency Vega under the 
contract no. 2/0058/26 and the project National competence centre for high performance 
computing within the Operational programme Integrated infrastructure 
(project code: 311070AKF2).

\end{document}